\documentclass[english]{article}
\usepackage[latin9]{inputenc}
\usepackage{geometry}
\usepackage{amsmath}
\usepackage{amssymb}

\makeatletter
\newcommand{\lyxaddress}[1]{
\par {\raggedright #1
\vspace{1.4em}
\noindent\par}
}

\makeatother

\usepackage{babel}
\begin{document}

\title{On the gravitomagnetic clock effect in quantum mechanics}

\author{S. B. Faruque%
\thanks{Corresponding Author: Email: awsbf62@yahoo.com, %
}, M. M. H. Chayon and Md. Moniruzzaman%
\thanks{Permanent Address: Department of Physics, Mawlana Bhashani Science
and Technology University, Santosh, Tangail-1902, Bangladesh%
}}

\maketitle

\lyxaddress{Department of Physics, Shahjalal University of Science and Technology,
Sylhet 3114, Bangladesh}
\begin{abstract}
In this article, we discuss how to carryover gravitomagnetic clock
effect from classical general relativity to quantum theory and how
to calculate this effect in quantum mechanics. Our calculation is
valid for semi-classical regime and can be considered as the first
step towards a complete gravitomagnetic quantum theory. We also show
the analogy between energy levels that corresponds to the clock effect.
In fact, it is argued that in quantum mechanics clock effect arises
as energy level splitting in gravitomagnetic field.
\end{abstract}

\section{Introduction}

General relativity predicts that two freely counter rotating test
particles in the exterior field of a central rotating mass take different
periods of time to complete the same full orbit; this time difference,
first discovered by Cohen and Mashhoon {[}1{]} in 1993, is termed
as gravitomagnetic clock effect. Since its first formulation, numerous
works have been devoted to elucidate its foundation {[}2-5{]} and
to derive methods to detect it via space borne experiments {[}6-12{]}.
Gravitomagnetic clock effect (GCE) has been extended by Faruque to
include spin of the test particle {[}13{]} and the spin-dependence
is elucidated further by Faruque {[}14{]} and by Bini et al{[}15{]}.
Mashhoon et al {[}16{]} shed more light on the spin dependence of
GCE in their work on extended spinning masses. Now, the GCE is one
of many gravitomagnetic effects such as, Lense-Thirring precession
of gyroscopes, gravitomagnetic time delay, spin-rotation coupling
etc., which are awaiting detection in astrophysical systems and also
are carrying over to quantum mechanics. Discovering gravitomagnetism
in quantum mechanics is a very recent physicist's drive which was
founded by Adler and Chen{[}17{]}. Adler and Chen {[}17{]} considered
Klein-Gordon equation expressed in generally covariant form and coupled
to electromagnetic field and found the analog of Lense-Thirring precession
in quantum system. Their work endows a spin-zero test mass in orbit
in a gravitomagnetic field with a gravitomagnetic moment $\overrightarrow{\mu}_{grav}$=$-\frac{1}{2}$$\overrightarrow{L}$,
where $\overrightarrow{L}$ is the orbital angular momentum. Finding
inspiration from Adler and Chen, we have tried to carryover GCE from
classical general relativity to quantum mechanics without actually
solving any equation but using the most usual prescription of quantum
to classical correspondence as formulated by Bohr. In Section 2, we
shall present our main derivation of GCE in quantum system. In Section
3, we shall discuss energy considerations. In Section 4, we will summarize
our work.

\section{GCE in quantum system}

To carry the classical GCE over to quantum mechanics, we shall not
try a fully relativistic and fully quantum mechanical formulation
of the physical problem that resembles the classical system which
shows GCE. Rather, we shall carry the physical problem showing GCE
conceptually to a quantum description. We shall use Bohr's correspondence
principle to convert classical angular frequencies to quantum energies.
To begin with, we note that the formula for orbital angular frequency
of a test particle in circular orbit around a central Kerr source
of spin parameter $a$ $=\frac{S}{Mc^{2}}$ , $S$ and $M$ being
the spin angular momentum and mass of the central rotating body, respectively,
 and c is the speed of light in vacuum, is, for corotating orbit,
\begin{equation}
\frac{1}{w}=a+\frac{1}{w_{k}}
\end{equation}
 And that for counter rotating orbit is
\begin{equation}
\frac{1}{w}=a-\frac{1}{w_{k}}
\end{equation}
where $w_{k}$ is the Keplerian angular frequency of orbital motion.
We now divide Eqs.(1) and (2) by $\hbar$ to find the semiclassical
energy of a test particle in states of energies $E_{\pm}$and of z-component
of orbital angular momentum $L_{z}=\pm$$\hbar$. These energies are
\begin{equation}
\frac{1}{E_{\pm}}=\frac{a}{\hbar}\pm\frac{1}{\hbar w_{k}}
\end{equation}
In Eq.(3), $E_{+}$corresponds to a state of z-component of orbital
angular momentum $L_{z}=+\hbar$ and $E_{-}$corresponds to a state
of z-component of orbital angular momentum $L_{z}=-\hbar$. This extension
of Eqs.(1) and (2), which are classical, to Eq.(3), which is semiclassical,
is obvious from the standpoint of Bohr's correspondence principle
. The carrying over of classical motion in corotating (counterrotating)
state to quantum physics and finding out the corresponding (gedanken)
time period for one complete revolution in orbit is accomplished by
demanding that the quantum state undergoes the transformation of the
azimuth angle $\phi\rightarrow\phi+2\pi$ and the time undergoes the
transformation $t\rightarrow t+T$, where T is time period for a full
cycle, which in later calculations will be written as $T_{\pm}$to
refer to co- and counter rotating orbits which are quantum mechanically
$L_{z}=\pm\hbar$ orbits. Before carrying out these operations, we
note that the only relevant wavefunction of the test particle in the
problem under consideration is an eigenstate of z-component of orbital
angular momentum and energy. The relevant states can be written as
\begin{equation}
\psi_{1}(t,\phi)=A\exp(i\phi)\exp(-\frac{i}{\hbar}E_{+}t)
\end{equation}
and 
\begin{equation}
\psi_{2}(t,\phi)=A\exp(-i\phi)\exp(-\frac{i}{\hbar}E_{-}t)
\end{equation}
In Eq.(4), the state is of energy $E_{+}$and $L_{z}=+\hbar$ and
in Eq.(5), the state is of energy $E_{-}$ and $L_{z}=-\hbar$. The
operations mentioned above produces the states $\psi_{1}(t+T_{+},\phi+2\pi)$
and $\psi_{2}(t+T_{-},\phi+2\pi)$. We appropriately then demand single-valuedness
of the quantum wavefunctions and thereby impose
\begin{equation}
\psi(t,\phi)=\psi(t+T_{\pm},\phi+2\pi)
\end{equation}
One can follow the mathematical steps when implementing the condition
(6) on the functions (4) and (5) and those generated from (4) and
(5) using the transformations just mentioned. By this way one automatically
finds 
\begin{equation}
T_{+}=\frac{2\pi\hbar}{E_{+}}
\end{equation}
and 
\begin{equation}
T_{-}=-\frac{2\pi\hbar}{E_{-}}
\end{equation}
Using Eqs.(7) and (8), one finds that

\begin{equation}
T_{+}-T_{-}=4\pi a
\end{equation}
which is the established classical GCE. We have thus carried the classical
GCE over to quantum mechanical GCE. Both of these are numerically
exactly the same. However, one can argue that the states $E_{+}$and
$E_{-}$with $L_{z}=\pm\hbar$ are different states and the GCE calculated
using them in quantum settings do not mimic the classical picture
and so cannot be called the quantum GCE. This is in one sense right,
but what is most important in this regard to realize is that what
in general relativity is termed as GCE, appears in quantum mechanics
as splitting of energy levels through gravitomagnetism. The gravitomagnetic
field of the central spinning body ( the Kerr source) produces a magnetic
field in which the gravitomagnetic moment of the test particle finds
itself and the energy of the test particle is different in the two
$L_{z}=\pm\hbar$ states according gravitomagnetic potential energy
$-\overrightarrow{\mu}_{grav}.\,\overrightarrow{B}$. Thus a single
energy state splits into two ( not three, the $L_{z}=0$ state is
missing) states of different total energy. Thus, we get hyperfine
splitting. In the next section , we discuss this feature.

\section{Energy considerations}

As shown in {[}17{]}, gravitomagnetic moment of a spin -zero test
mass in orbit in a gravitomagnetic feild is given by 
\begin{equation}
\overrightarrow{\mu}_{grav}=-\frac{1}{2}\overrightarrow{L}
\end{equation}
Now, let us consider a spining ball of mass $M$ and radius$R,$ spinning
with angular frequency $\overrightarrow{\omega}$. On the equarorial
plane the gravitomagnetic feild $\overrightarrow{B}$ produced by
the ball is given by {[}18{]}
\begin{equation}
\overrightarrow{B}=+\frac{4}{5}\frac{GR^{2}M\overrightarrow{\omega}}{c^{2}r^{3}}
\end{equation}
This can be rewritten as 
\begin{equation}
\overrightarrow{B}=\frac{2GI\overrightarrow{\omega}}{c^{2}r^{3}}=\frac{2\overrightarrow{J}}{Mc^{2}}\omega_{k}^{2}
\end{equation}
where the moment of the inertia of the ball is $I=\frac{2}{5}MR^{2}$and
its spin angular momentum $\overrightarrow{J}=I\overrightarrow{\omega}$.
The angular frequency of the test particle in orbit is $\omega_{k}=\sqrt{\frac{GM}{r^{3}}}$
. 

As the system we are considering is that of a test mass either in
prograde orbit or in retrograde orbit, the test particle orbital angular
momentum along the spin axis is $L_{z}=+\hbar$ for the prograde orbit
and that is $L_{z}=-\hbar$ for the retrograde orbit.Thus, we find
the potential energy associted with the gravitomagnetic moment as
\begin{equation}
E_{B}=-\overrightarrow{\mu}_{grav}.\overrightarrow{B}\equiv\frac{\overrightarrow{L}.\overrightarrow{J}}{Mc^{2}}\omega_{k}^{2}=\begin{cases}
\begin{array}{cc}
\frac{\hbar\omega_{k}^{2}J}{Mc^{2}} & ,L_{z}=+\hbar\\
-\frac{\hbar\omega_{k}^{2}J}{Mc^{2}} & ,L_{z}=-\hbar
\end{array}\end{cases}
\end{equation}
The total energy in $L_{z}=+\hbar$ and $L_{z}=-\hbar$ states are
found by adding the principle energy of the test mass which is $-\hbar\omega_{k}$,
the Keplerian energy. So, we have 
\begin{equation}
\mathbf{E_{+}=}-\hbar\omega_{k}+\frac{\hbar\omega_{k}^{2}J}{Mc^{2}}
\end{equation}
\begin{equation}
\mathbf{E_{-}=}-\hbar\omega_{k}-\frac{\hbar\omega_{k}^{2}J}{Mc^{2}}
\end{equation}
We can rearrange Eqs. (14) and (15) as 
\begin{equation}
\mathbf{E_{+}=}-\hbar\left(\omega_{k}-\frac{\omega_{k}^{2}J}{Mc^{2}}\right)
\end{equation}
\begin{equation}
\mathbf{E_{-}=}-\hbar\left(\omega_{k}+\frac{\omega_{k}^{2}J}{Mc^{2}}\right)
\end{equation}
We can identify the orbital frequency in $E_{+}$state as
\begin{equation}
\omega_{+}=\omega_{k}-\frac{\omega_{k}^{2}J}{Mc^{2}}
\end{equation}
and in $E_{-}$state as 
\begin{equation}
\omega_{-}=\omega_{k}+\frac{\omega_{k}^{2}J}{Mc^{2}}
\end{equation}
The period of revolution in $E_{+}$state is 
\[
T_{+}=\frac{2\pi}{\omega_{+}}=\frac{2\pi}{\omega_{k}-\frac{J\omega_{k}^{2}}{Mc^{2}}}
\]

\[
=\frac{2\pi}{\omega_{k}}\left(1+\frac{J\omega_{k}}{Mc^{2}}\right)
\]
\begin{equation}
=\frac{2\pi}{\omega_{k}}+\frac{2\pi J}{Mc^{2}}
\end{equation}
The period of revolution in $E_{-}$state is
\[
T_{-}=\frac{2\pi}{\omega_{-}}=\frac{2\pi}{\omega_{k}+\frac{J\omega_{k}^{2}}{Mc^{2}}}
\]

\[
=\frac{2\pi}{\omega_{k}}\left(1-\frac{J\omega_{k}}{Mc^{2}}\right)
\]
\begin{equation}
=\frac{2\pi}{\omega_{k}}-\frac{2\pi J}{Mc^{2}}
\end{equation}
Subtracting Eq. (21) from Eq. (20), we obtain 
\begin{equation}
T_{+}-T_{-}=\frac{4\pi J}{Mc^{2}}
\end{equation}
which is the clock effect (Eq. (9) with$a=\frac{J}{Mc^{2}}$ ).

Therefore, we have become able to show that the quantum mechanical
energy of a gravitomagnetic dipole of moment $\overrightarrow{\mu}_{grav}=-\frac{1}{2}\overrightarrow{L}$
in gravitomagnetic field $\overrightarrow{B}$ , produced by a rotating
mass, added with the principal Keplerian quantum of energy $-\hbar\omega_{k}$
, gives two states $E_{+}$and $E_{-}$given in Eqs. (14) and (15)
. The angular frequences in these two states are given in Eqs. in
(18) and (19). The corresponding periods are $T_{+}$and $T_{-}$
which are same as the classical periods giving rise to the classical
gravitomagnetic clocl effect.

However, the states given in Eq.(3) are not same as those given in
Eqs. (14) and (15). There is a fundamental differance in the derivation
followed in this section and that in the previous section. Energy
considerations followed in this section are more plausible than that
followed in the previous section. To adjust the two approaches we
note that the energy $E_{+}$given by Eq.(3) is not a bound state,
but it can be made to correspond to a bound state by putting a $\left(-\right)$in
front. On the other hand the energy $E_{-}$ given by Eq. (3) in that
of a bound state and it is correct as it is. That is $E_{+}\left(E_{-}\right)$given
by Eq. (3) becomses equivalent to those given by Eq. (14) and Eq.
(15) only when we perform the adjustment just mentioned.

Finally, we note that the derivation with $E_{\pm}$given by Eq. (3)
followed in section 2 is quantum theoretically true in the sense that
a positive time should correspond to a positive energy which is true
for both $T_{+}$and $T_{-}$ given in Eqs. (7) and (8). Thus, keeping
in mind these artifacts of a semi-classical theory, we understand
that gravitomagnetic clock effect indeed has a counterpart in quantum
theory appearing through energy considerations in gravitomagnetic
field.

\section{Summary}

In this article, we have discussed the way to carry the classical
gravitomagnetic clock effect in Kerr field and in GEM field over to
quantum physics. We have used Bohr's correspondence principle to convert
classical orbital angular frequencies to semiclassical energies. Thus,
we have found the corotating and counterrotating orbits in Kerr field
to correspond to two different states having different energy and
different angular momentum (orbital z-component). When we apply appropriate
operations to generate states that help to calculate time periods
in a quantum sense and demand single- valuedness of wavefunctions,
we automatically find a quantum analog of GCE which is numerically
the same as classical GCE. We have thus shown a particular mechanism
for discussing GCE in a quantum system. 

We also have discussed actual energy considerations involved with
the system. A particle with gravitomagnetic moment in a gravitomagnetic
field possesses potential energy. When this energy is added with the
principlal Keplerian energy we atumatically find the classical GCE
to appear in the quantum system. Note that if the quantum condition
of $L_{z}=\pm\hbar$ is not applied but $L$ is left classical then
the clock effect would not correspond to what actually is called GCE.
Moreover,only micro black holes with quantized spin are candidates
of sources of GEM field in which micro-particles with Compton radius
below Planck length can show quantum GCE.

Quantum GCE in ordinary quantum systems is so insignifical in magnitude
that it might not be measurable in near and far future. However, as
long as a particular type of classical GCE depends on only intrinsic
properties of orbiter and central rotator, such as spin and mass,
the quantum GCE will have the same formula. The numeral insignificance
of quantum GCE does not mean that it is not an element of reality.
Because reality basically is of quantum nature. Whatever physical
phenomena appear in classical physics should have a corresponding
quantum phenomena. And quantum phenomena actually are in the root
of classical phenomena. Hence, what we have discussed in this article,
that is quantum GCE, cannot be denied of existence if classical GCE
appears as reality.

\end{document}